# $MgB_2$ tunnel junctions with native or thermal oxide barriers

R. K. Singh, R. Gandikota, J. Kim, N. Newman[a], and J. M. Rowell
Department of Chemical and Materials Engineering, Arizona State University, Tempe,
AZ 85287-6006

$MgB_2$ tunnel junctions ($MgB_2$/barrier/$MgB_2$) were fabricated using a native oxide grown on the bottom $MgB_2$ film as the tunnel barrier. Such barriers therefore survive the deposition of the second electrode at 300°C, even over junction areas of ~1 mm$^2$. Studies of such junctions, and those of the type $MgB_2$/native or thermal oxide/metal (Pb, Au, or Ag) show that tunnel barriers grown on $MgB_2$ exhibit a wide range of barrier heights and widths.

[a] electronic mail: Nathan.Newman@asu.edu

Magnesium diboride (MgB$_2$) has the potential to be used in Josephson junction devices operating at 20 to 25 K because it has a transition temperature of 40 K,[1] an anisotropy of 5-7,[2] and a coherence length of ~5 nm.[3] Several groups have already demonstrated tunnel junctions with both lower T$_c$ superconductors[4-7] and with MgB$_2$ as the top electrode[8-10] using different barrier layers such as AlN, MgO, Al$_2$O$_3$, and the thermal oxide. The results indicate that key issues for fabricating tunnel junctions are the uniform coverage of the base electrode by these barriers, as well as the stability of these barriers when the top MgB$_2$ electrode is deposited at temperatures ≥300°C. In junctions with AlN barriers[9] the data suggested that a thermal oxide must be part of the barrier, and that stimulated our work. The utility of the native oxide is well known in the popular Nb/Al–Al oxide/Nb trilayer junction technology. In this paper, we demonstrate MgB$_2$ junctions with oxide barriers which exhibit good tunneling characteristics. The fact that fairly low leakage tunneling characteristics are found in junctions as large as 1 mm$^2$ demonstrates that the native oxide covers the base MgB$_2$ electrodes without pinholes. Our junctions, including those with both electrodes being MgB$_2$, exhibit superconducting energy gaps to over 30 K. In these junctions we see only the smaller π gap as might be expected given that the c-axis is the only orientation found when observable in X-ray diffraction and Transmission Electron Microscopy measurements. Conductance-voltage (dI/dV-V or G-V) measurements at voltages >~100mV indicate that the barrier height/thickness is significantly lower/larger than that reported by Schneider et al.[11] We suggest that the surface chemistry of the MgB$_2$, and oxidation conditions, can cause barrier properties to be so widely different.

The MgB$_2$ films were deposited at 300 ± 2 °C in an ultra-high vacuum MBE system with a base pressure of ~5×10$^{-10}$ Torr. The system pressure reached as high as 10$^{-6}$ Torr during deposition. Further growth details have been published elsewhere.[12,13] A 1 to 4 mm wide and 1500 Å thick strip of Mg-rich magnesium diboride (~ $Mg_{1.1}B_2$) was grown on (0001) sapphire substrate using a metal shadow mask. The reason for our choice of stoichiometry has been addressed later. The thickness and chemical composition were determined using Rutherford backscattering spectrometry. After the film had cooled to room temperature in the residual gases, the deposition system was vented to atmospheric pressure with nitrogen. The native oxide and thermal oxide barrier layers were formed by exposing the $MgB_2$ strip to ambient atmosphere at room temperature and 160 °C (on a hot plate), respectively, for one hour. However, we can not rule out the possibility that the barrier forms in the residual gases before venting. The edges of the $MgB_2$ strip were coated with Duco cement after oxidation to minimize edge effects. In the case of all-$MgB_2$ junctions, the cement was not used. A crossed-geometry junction was fabricated by depositing top $MgB_2$ (~1000 Å thick), or by thermally evaporating a metal (Pb, Au, or Ag) as four 1 mm wide strips aligned perpendicular to the bottom electrode strip. The performance of the tunneling barrier layer was also evaluated by defining $MgB_2$/native oxide/$MgB_2$ trilayer junctions using photolithography and etching ($BCl_3$ plasma). In this case, a 200 nm thick $SiO_2$ film was used as an insulation layer. Four-terminal current-voltage measurements were made with a Quantum Design Physical Property Measurement System and a dipping probe using direct contacts in the shadow mask junctions and Ti/Au contacts in the lithographic junctions. G-V plots were obtained by differentiating the measured I-V characteristics.

Figure 1 shows the normalized conductance plot of an all-$MgB_2$ tunnel junction (1 mm×1mm) with a native oxide barrier, defined using shadow masks. The 3.4 mV peak corresponds to the sum of the superconducting energy gaps of the top and bottom $MgB_2$ electrodes. The top electrode has a $T_c$ (zero resistance) of only 22 K, compared to 34 K in the bottom electrode (see inset of Fig. 1). Note the very broad transition of the top film, compared to the sharp one of the bottom film. High quality MgB$_2$ has a transition temperature of 39 K, and a π-gap energy (Δ$^π$) of 2.3 mV,[14] thus the ratio, 2Δ$^π$(0)/k$_B$T$_c$ is 1.36. If this ratio does not change with T$_c$, then the gap values of the two MgB$_2$ electrodes can be estimated to be 2.0 mV for the bottom electrode (T$_c$=34 K) and 1.3 mV for the top MgB$_2$ electrode (T$_c$=22 K). Thus, the sum of the two gap values (2.0 mV+1.3 mV) is close to the peak location at 3.4 mV in the tunneling conductance curve of Fig. 1. This scaling of Δ$^π$ with electrode T$_c$ implies that the gap suppression is not caused by degraded material adjacent to the barrier.

Figure 2(a) shows the conductance-voltage characteristics of a 100 μm×100 μm all-MgB$_2$ junction defined by photolithography and etching with a sum gap of 4.3 mV (at 4.2 K), implying that the T$_c$ of both films in this junction must at least be 34 K. With the small mesa-like geometry of lithographic trilayer junctions, we could not measure T$_c$ of the individual electrodes. The junction, in spite of the high T$_c$s, exhibited poor tunneling and sub-gap characteristics (R$_{SG}$=15.6Ω & R$_N$=14.3Ω) and no supercurrent was observed [see inset of Fig. 2(b)]. Also observed is a conductance peak at zero bias that grows in magnitude as the temperature is raised. This is presumably caused by quasiparticles thermally excited across the π-gap (~2.3 mV) at higher temperatures. To date, the reproducibility of the all-MgB$_2$ junctions described above has been low.

The two junctions, defined by shadow masks and by photolithography and etching, have R$_N$A values of 6.25×10$^{-1}$ and 1.4×10$^{-3}$ Ω.cm$^2$ respectively. The lack of supercurrents in the latter junction is perhaps surprising as all-MgB$_2$ junctions with AlO$_x$ barriers having similar R$_N$A values, reported by Ueda et al.,[10] show supercurrents. It should however be noted that in those junctions, I$_c$ was relatively small at 4.2 K, and much larger below 1 K. In our case, noise in the unshielded dipping probe/PPMS system might be depressing $I_c$ at our higher measurement temperatures ($\geq$4.2 K). The temperature dependence of the sum gap [Fig. 2(b)] shows superconducting energy gaps to over 30 K and fits closely to 2$\Delta$(T) evaluated using the expression[15] $\Delta(T)=\Delta(0)*[1-(T/T_c)^p]^{0.5}$ for $\Delta(0)$=2.2 mV, $T_c$=36.5 K, and *p*=2.

From our work on MgB$_2$/native oxide/Pb junctions with Mg-rich MgB$_2$ electrodes we found that 25 out of a series of 27 1 mm$^2$ junctions showed good gap characteristics, although the reproducibility of $R_N$ (~5×10$^2$ to 10$^6$ $\Omega$) was poor. To exclude the possibility that these characteristics arise as a result of oxidation at the surface of the top Pb electrode, studies were performed on junctions with Au or Ag top electrode. They also show similar tunneling characteristics. We also evaluated junctions with stoichiometric MgB$_2$ films with lower resistivities and higher transition temperatures. However, in junctions made with such stoichiometric MgB$_2$ electrodes, the yield of good characteristics was reduced. This suggests that the junction oxidation process is less reliable on stoichiometric MgB$_2$ films. The I-V characteristics of all our tunnel junctions show sub gap leakage currents, with sub gap conductance ($\sigma_{SG}$) typically varying from 10 % to over 90 % of the normal junction conductance ($\sigma_N$).

The oxidation of MgB$_2$ is more complex than that of, say, Al or Mg alone, as is used in both superconducting and magnetic junctions. The oxide layer can be MgO, BO$_x$, and/or MgB$_x$O$_y$. Photoemission measurements of MgB$_2$ films after exposure to air show large amounts of carbon and oxygen, but relatively little MgO and BO$_x$. The presence of nitrides of boron was not detected.[16,17] The surface stoichiometry of MgB$_2$ films depends on the film composition and can be altered by depositing a thin layer of Mg or boron on the surface. This can ultimately determine the barrier behavior.

It has been well established that tunneling currents are ohmic at low voltages and depend exponentially on voltage at high voltages.[18] At moderate voltages (>~100 mV) the barrier height and thickness can be inferred from the conductance versus voltage dependence. This characterization technique has been used recently by Schneider et al.[11] for $MgB_2$/oxide/In junctions, and they find the barrier height ($\varphi$) and thickness (s) to be 1.6 eV and 1.5 nm respectively, which are parameters very similar to those of Al oxide junctions.[18] All the native and thermal oxide barriers that we have used to date have very different barrier heights and widths. Fig. 3(a) shows the G-V dependences for 5 different junctions, while the insets in Fig. 3(b) show the I-V characteristics of one of these junctions at low and high bias respectively. That these junctions have low barrier heights can immediately be inferred from the fact that the bias required to double the minimum conductance ($\alpha$, usually at V=0 or very close to it) is typically 100 mV or less, whereas in Al oxide junctions it is ~550 mV. For all our junctions we find that $\varphi$ ranges from 0.11 to 0.33 eV, and s from 4.1 to 5.4 nm. In Fig. 3(b), it is surprising that the $V^2$ dependence of G persists to voltages well above the barrier heights calculated from $\alpha$ and the voltage at $2\alpha$. The conductance is expected to increase more rapidly than $V^2$ at higher voltages. So we calculated G(V) by evaluating the full expression (equation 27 of Ref. 19) using the obtained barrier height (0.15 eV) and width (5.4 nm) for one of the junctions. Indeed, G increased much more rapidly than the data at voltages comparable to the barrier height. Further studies to understand this unexpected high-voltage transport are ongoing.

These barrier heights that we have obtained for the native and thermal oxides are surprisingly low, although we note that the reported values of potential height of MgO barriers in magnetic junctions are from 0.39 eV to 3.7 eV.[20-22] We have preliminary evidence that boron plays a role in our junctions. We prepared two $MgB_2$ films with ~5 and ~10 nm of boron on the surfaces and made junctions MgB$_2$/B/oxide/Pb in the usual way. The junction with 10 nm of boron was of extremely high resistance and had a strongly non-linear I-V at high voltages (~0.4 volts). The junction with 5 nm of boron showed a poorly defined sum gap ($\Delta_{Pb}+\Delta_{MgB2}$), and G depended strongly on voltage, in fact the minimum G (at V=0) was doubled at only 8 mV. It is not certain at present whether the transport across these barriers on boron is by tunneling, or some kind of hopping transport. Zeller et al.[23] have shown that features associated with the gap can still be observed even in the case of hopping transport through the barrier.

To summarize, we have demonstrated MgB$_2$/native oxide/MgB$_2$ junctions that showed tunneling characteristics using two different junction definition routes viz., cross-geometry junctions using shadow masks and junctions formed using photolithography and etching. The sum gap (of π-gaps), obtained with the latter route, was 4.3 mV and this gap remains non-zero for temperatures above 30 K. These results also demonstrate the pinhole-free coverage of the native oxide barrier layer over areas larger than 1 mm$^2$ and the survival of the same at temperatures at least up to 300°C, thus exhibiting great potential for use in trilayer junction technology. The barrier width and height that we obtained are very different from those reported for thermal oxide junctions reported by Schneider et al.[11] This suggests that native or thermal oxide barrier properties are sensitive to both surface stoichiometry and oxidation conditions.

The authors thank Deborah Van Vechten for her encouragement and support, as well as Hongxue Liu for assistance in the PPMS measurement. We would like to acknowledge the Office of Naval Research for support of this work under contract number N00014-05-1-0105.

[1] J. Nagamatsu, N. Nakagawa, T. Muranaka, Y. Zenitani, and J. Akimitsu, Nature (London) **410**, 63 (2001).

[2] S. L. Bud'ko, V. G. Kogan, and P. C. Canfield, Phys. Rev. B **64**, 180506(R) (2001).

[3] D. K. Finnemore, J. E. Ostenson, S. L. Bud'ko, G. Lapertot, and P. C. Canfield, Phys. Rev. Lett. **86**, 2420 (2001).

[4] G. Carapella, N. Martucciello, G. Costabile, C. Ferdeghini, V. Ferrando, and G. Grassano, Appl. Phys. Lett. **80**, 2949 (2002).

[5] T. H. Kim and J. S. Moodera, Appl. Phys. Lett. **85**, 434 (2004).

[6] J.Geerk, R. Schneider, G. Linker, A.G. Zaitsev, R. Heid, K.-P. Bohnen, and H. v. Lohneysen, Phys. Rev. Lett. **94**, 227005 (2005).

[7] A. Saito, A. Kawakami, H. Shimakage, H. Terai, and Z. Wang, J. Appl. Phys. **92**, 7369 (2002).

[8] D. Mijatovic, A. Brinkman, I. Oomen, G. Rijnders, H. Hilgenkamp, H. Rogalla, and D. H. A. Blank, Appl. Phys. Lett. **80**, 2141 (2002).

[9] H. Shimakage, K. Tsujimoto, Z. Wang, and M. Tonouchi, Appl. Phys. Lett. **86**, 072512 (2005).

[10] K. Ueda, S. Saito, K. Semba, T. Makimoto, and M. Naito, Appl. Phys. Lett. **86**, 172502 (2005).

[11] R. Schneider, J. Geerk, F. Ratzel, G. Linker, and A.G. Zaitsev, Appl. Phys. Lett. **85**, 5290 (2004).

[12] J. Kim, R. K. Singh, N. Newman, and J. M. Rowell, IEEE Trans. Appl. Supercond. **13**, 3238 (2003).

[13] J. Kim, R. K. Singh, N. Newman, J. M. Rowell, L. Gu, and D. J. Smith, J. Crystal. Growth **270**, 107 (2004).

[14] M. Iavarone, G. Karapetrov, A.E. Koshelev, W.K. Kwok, G.W. Crabtree, D.G. Hinks, W.N. Kang, E-M Choi, H.J. Kim, H-J Kim, and S.I. Lee, Phys. Rev. Lett. **89**, 187002 (2002).

[15] H.J. Choi, D. Roundy, H. Sun, M.L. Cohen, and S.G. Louie, Nature (London) **418**, 758 (2002).

[16] B. Moeckly and R. Buhrman (unpublished).

[17] R. Gandikota, R.K. Singh, J.M. Rowell, N. Newman (unpublished).

[18] J.M. Rowell, in *Tunneling Phenomena in Solid*, edited by E. Burstein and S. Lundqvist (Plenum, New York, 1969), p.385.

[19] J.G. Simmons, J. Appl. Phys. **34**, 1793 (1963).

[20] S. Yuasa, T. Nagahama, A. Fukushima, Y. Suzuki, and K. Ando, Nat. Mater. **3**, 868 (2004).

[21] S.S.P. Parkin, C. Kaiser. A. Panchula, P.M. Rice, B. Hughes, M. Samant, and S.H. Yang, Nat. Mater. **3**, 862 (2004).

[22] W. Wulfhekel, M. Klaua, D. Ullmann, F. Zavaliche, J. Kirschner, R. Urban, T. Monchesky, and B. Heinrich, Appl. Phys. Lett. **78**, 509 (2001).

[23] H.R. Zeller and I. Giaever, Phys. Rev. **181**, 789 (1969).

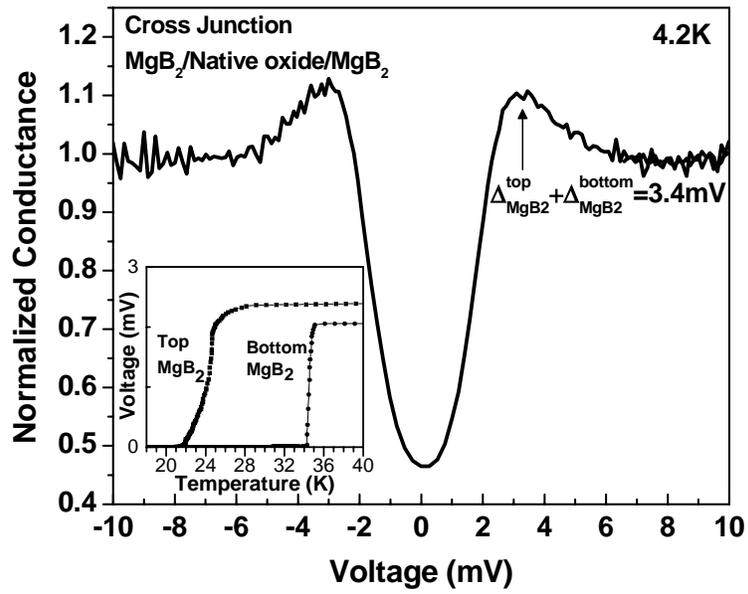

**Figure 1 – Singh et al.**

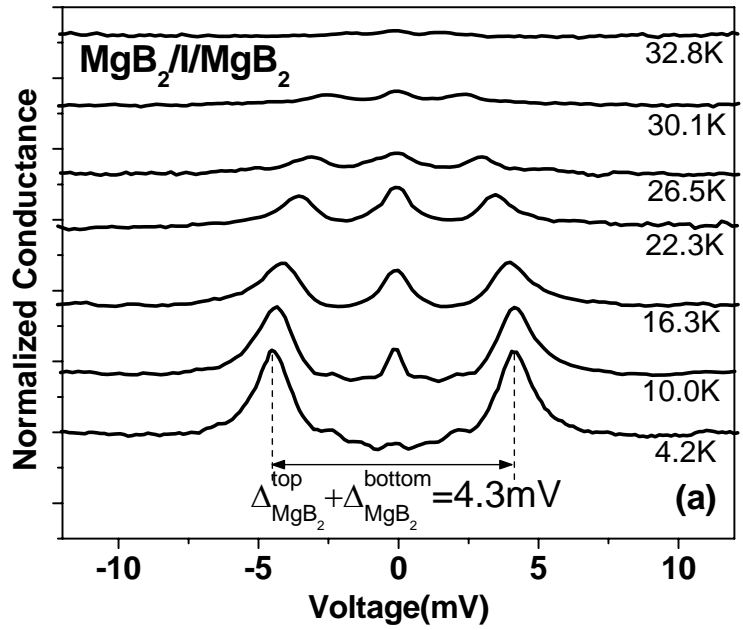

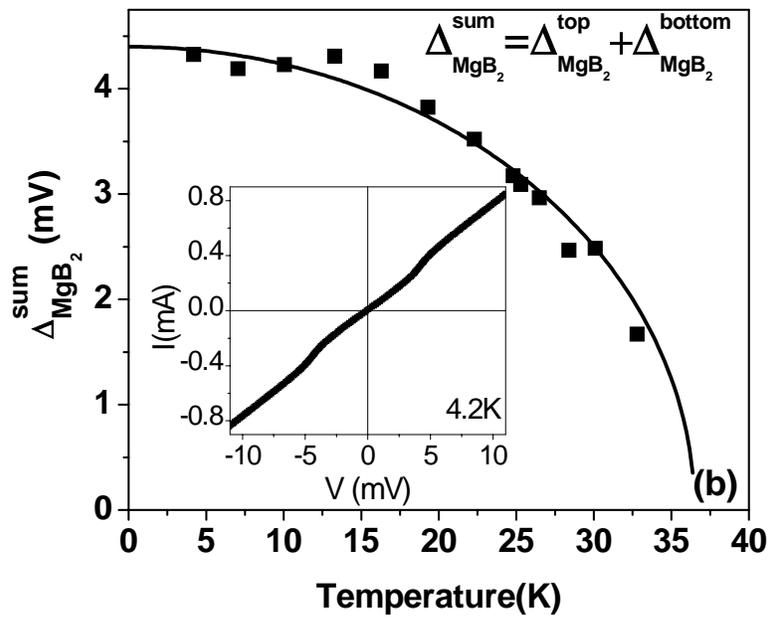

**Figure 2- Singh et al.**

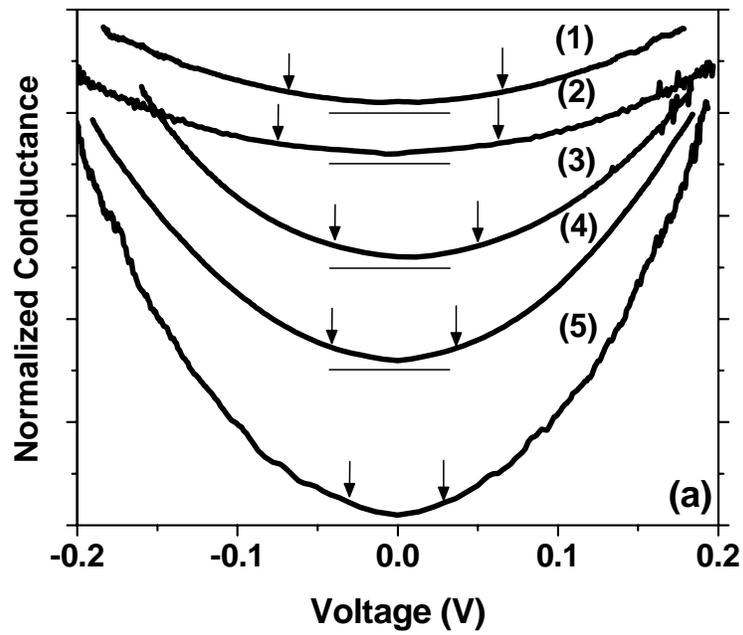
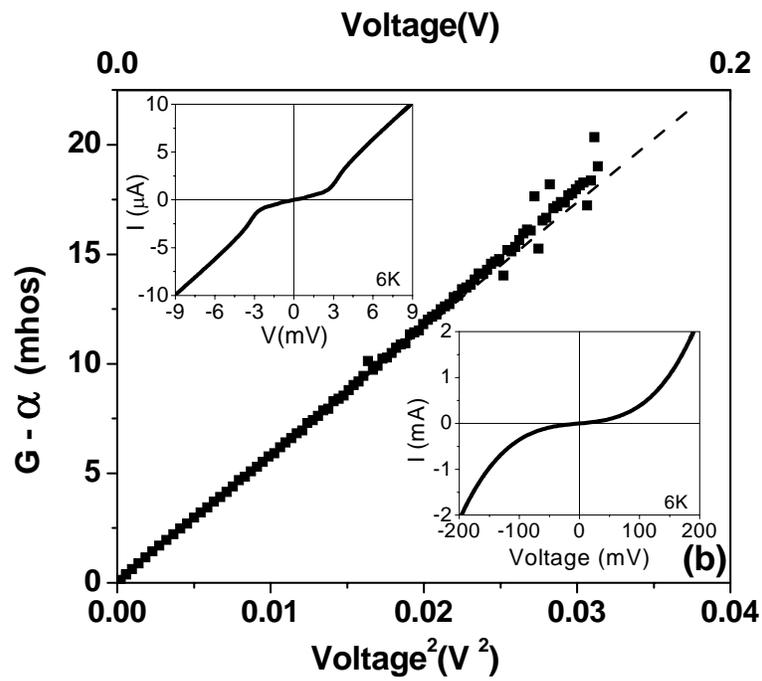

**Figure 3 – Singh et al.**

FIGURE CAPTIONS

Figure 1. Normalized conductance plot measured at 4.2 K for an $MgB_2$/native oxide/$MgB_2$ junction. Inset shows the transition temperatures of top and bottom $MgB_2$ electrodes.

Figure 2. (a) Normalized conductance vs. Voltage at select temperatures for an $MgB_2$/native oxide/$MgB_2$ junction defined using lithography and etching. (b) Variation of the sum gap of the two electrodes as a function of temperature: discrete points were experimentally obtained while the continuous line was obtained by theoretical evaluation. Inset in Fig. 2(b) shows the I-V characteristics at low bias.

Figure 3. (a) Normalized conductance versus voltage plot for $MgB_2$/oxide/Pb junctions with native oxide barriers (curves marked 2 and 3) and with thermal oxide barriers (curves marked 1, 4, and 5). On each of the curves, the arrows point to the voltages at which the conductance is twice that at zero bias. The curves have been shifted vertically and the horizontal line beneath each curve is the zero reference. All G-V curves were obtained at temperatures between 4.2 and 6.5 K. (b) (G-$\alpha$) versus $V^2$ plot for the $MgB_2$/native oxide/Pb junction referred as (3) in Fig. 3(a). The two insets in Fig. 3(b) show the I-V characteristics at low and high bias respectively, for the junction marked (4) in Fig. 3(a).